# Data Conditioning for Subsurface Models with Single-Image Generative Adversarial Network (SinGAN)

Lei Liu[a], Eduardo Maldonado-Cruz[b], Honggeun Jo[c], Maša Prodanović[a] and Michael J. Pyrcz[a,d]

[a]Hildebrand Department of Petroleum and Geosystems Engineering, The University of Texas at Austin, Austin, TX, USA

[b]Chevron Energy Technology Company, Houston, TX 77002, USA

[c]Energy Resources Engineering, Inha University, Incheon, Korea

[d]Department of Planetary and Earth Sciences, The University of Texas at Austin, Austin, TX, USA

Keywords: data conditioning; minimum acceptance checks; subsurface models; GAN

Abstract

The characterization of subsurface models relies on the accuracy of subsurface models which request integrating a large number of information across different sources through model conditioning, such as data conditioning and geological concepts conditioning. Conventional geostatistical models have a trade-off between honoring geological conditioning (i.e., qualitative geological concepts) and data conditioning (i.e., quantitative static data and dynamic data). To resolve this limit, generative AI methods, such as Generative adversarial network (GAN), have been widely applied for subsurface modeling due to their ability to reproduce complex geological patterns. However, the current practices of data conditioning in GANs conduct quality assessment through ocular inspection to check model plausibility or some preliminary quantitative analysis of the distribution of property of interests. We propose the generative AI realization minimum acceptance criteria for data conditioning, demonstrated with single image GAN. Our conditioning checks include static small-scale local and large-scale exhaustive data conditioning checks, local uncertainty, and spatial nonstationarity reproduction checks. We also check conditioning to geological concepts through multiscale spatial distribution, the number of connected geobodies, the spatial continuity check, and the model facies proportion reproduction check. Our proposed workflow provides guidance on the conditioning of deep learning methods for subsurface modeling and enhanced model conditioning checking essential for applying these models to support uncertainty characterization and decision making.

1. Introduction

Subsurface modeling is a crucial process in the field of reservoir engineering, groundwater management, mineral mining, and hydrocarbon exploration. Accurate subsurface models are essential for uncertainty characterization and decision-making. The characterization and reliability of these models require integrating a maximum amount of information across different scales and sources through model conditioning. The integrated conditioning includes geological concepts conditioning (e.g., geometries from interpretations of analog outcrops, hydrogeology data from aquifer property), static small-scale local measures data conditioning (e.g., local feature values from interpreted well log data), and static large-scale exhaustive data conditioning (e.g., horizons, faults, reservoir extents, and feature trends from interpretation of seismic attribute such as acoustic impedance), and dynamic grade-forecasting data conditioning (e.g., historical production and injection) and dynamic time-lapse (or 4D) data conditioning (e.g., groundwater flow rates, seismic flow barriers and conduits) (Bosch et al., 2009; Deutsch, 1993; Feyen & Gorelick, 2004; Lødøen & Omre, 2008; Pyrcz & Deutsch, 2014).

The traditional geostatistical methods are designed to integrate and reconcile diverse types of conditioning data and geological information through two-point semivariograms, multiple-point statistics (MPS), parameterized objects, and geometric rules (Michael et al., 2010; Pyrcz & Deutsch, 2014; Pyrcz et al., 2012, 2015; Strebelle & Journel, 2001). An accurate and realistic subsurface model is motivated by conditioning to not only exact static data (such as well data, groundwater data, and seismic interpretations), but also geological concepts (such as deposit element's geometries or stacking patterns, and sedimentary sequences). Yet, there is a trade-off between honoring geological conditioning (i.e., qualitative geological concepts) and data conditioning (i.e., quantitative static data and dynamic data) for geostatistical methods.

Semivariogram-based methods directly constrain the geostatistical models to the conditional data by assigning these conditional data to the model grids and then simulating other nodes following a random path. This cell-based method uses two-point variogram statistics to constrain the spatial continuity between the grid cells, thus having virtually unlimited ability to condition to petrophysical measures at wells and trend models from seismic data, but limited geological concept conditioning capability, i.e., an inability to model more complex spatial relationships (e.g., curvilinear structures) (Bertoncello et al., 2013; Pyrcz & Deutsch, 2014). The multiple-point-based algorithms follow the same steps as the variogram-based methods and simulate grid cells sequentially along a random path based on the search trees constructed from multiple-point statistics from the training image (TI) (Guardiano & Srivastava, 1993; Strebelle, 2002). MPS-based approaches retain the same ability to condition to petrophysical and seismic data and also demonstrates improved performance in geological concept conditioning through facies modeling with multiple points with the ability to model curvilinear and ordering relationships. Yet, MPS-based approaches, however, still exhibit suboptimal performance in effectively integrating detailed geological information from the training image, such as geometries and stacking sequences (Pyrcz et al., 2015).

The object-based methods randomly place the objects that represent critical elements of the subsurface (e.g., channels, levees, and lobes for constructing reservoir models) with appropriate shapes and dimensions to the model, thus conditioning to geological concepts based on geometry, such as the long-range curvilinear structures, but conditioning dense datasets can be problematic in objected-based methods with outcomes from significantly increased computational intensity and slow convergence up to failure to match petrophysical and seismic conditioning data. (Hauge et al., 2007; Michael et al., 2010; Skorstad et al., 1999; Strebelle & Remy, 2005). The rule-based method, a variant of object-based method, considers the concept of time (stratigraphic sequences) and condition to the hierarchical events during modeling, however conditioning to dense data may be challenging due to constrained rules between objects may further prevent conditioning in forward modeling (Pyrcz et al., 2012).

To resolve the limits in conventional geostatistical modeling, deep learning has been increasingly applied to subsurface modeling (Jo et al., 2020; Liu, Mehana, et al., 2024; Mabadeje & Pyrcz, 2024; Maldonado-Cruz & Pyrcz, 2022; Mosser et al., 2017). A common applied framework for realization generation is generative adversarial network (GAN), which consists of two neural networks, a generator and a discriminator. The generator generates artificial data while the discriminator discriminates the generated data and real data (Goodfellow et al., 2014). GAN has shown its advantages in characterizing and reproducing spatial relationships in subsurface modeling through various applications, such as porous media reconstruction (Mosser et al., 2017), channelized reservoir modeling (Liu et al., 2024), and rule-based lobe modeling (Jo et al., 2021). Different from GANs that require multiple TIs as training data that may be time-consuming due to TIs collection and increased computational burden, the development of single-image GAN (SinGAN) trains with a single image and generates realizations from the TI (Shaham et al., 2019). Liu et al. (2024) demonstrate the capability of applying SinGAN for channelized reservoir modeling, specifically SinGAN's ability in reproducing spatial nonstationarity. The introduction to the SinGAN architecture is available in Appendix 1.

With the GAN method, conditioning is integrated during model training; therefore, all subsequently generated models are conditioned post-training via post-optimization. One common method for conditioning during model training is to concatenate the conditioning into the network as part of the input vector to the generator. For example, Chan and Elsheikh (2019) perform spatial observation conditioning by stacking a second inference network along with the existing generator. Another method for conditioning during modeling training is to integrate the condition to the part of loss function during the training process, such as the local well observations conditioning through a content loss (Pan et al., 2022; Zhang et al., 2021) and measured data conditioning for pore-scale models through content and perceptual loss (Mosser et al., 2018). This direct conditioning can ensure the exact data reproduction at specified locations, however, typically requires retraining of the neural network with the addition of any new conditioning data.

Post-optimization (or post-processing) is another method for integrating conditioning into realizations. This approach involves iteratively updating an initial model to incorporate desired data conditioning, thus avoiding the need to retrain the neural network model from scratch which can be time-consuming and computationally intensive with the addition of new conditioning data (Song et al., 2022). For example, Dupont et al. (2018) integrate petrophysical measurements conditioning into realistic geological realizations through post-optimization. Jo et al. (2020) integrate static petrophysical conditioning well data to rule-based models generated from a trained GAN model.

Prior to their implementation for subsurface development appropriate decision-making, rigorous modeling checking, based on conditioning checks is best practice (Boisvert et al., 2010; Leuangthong et al., 2004). These modeling data conditioning minimum acceptance checks include some basic qualitative checks, such as the direct visualization comparison of generated conditional realizations and TIs to see if original patterns are correctly reproduced, or visualize them in the reduced-dimensional space to check the similarity using dimensionality reduction method such as multidimensional scaling visualization, or the comparisons with the realizations generated through different methods, such as GAN and SNESIM (Chan & Elsheikh, 2019; Dupont et al., 2018; Jo et al., 2020; Strebelle, 2000).

Other modeling data conditioning minimum acceptance checks are quantitative checks, such as the conditioning accuracy based on the content loss which measures the L1 distance between well locations of true image and generated conditional realizations, or the estimated probability that a realization is real using the discriminator (Chan & Elsheikh, 2019; Zhang et al., 2021).

Unsolved problems remain in the quality assessment of GAN-related conditioning workflow for subsurface modeling. While these workflows demonstrate the methods for generating data conditioning geological realizations, the conditioning assessment is either qualitative visualization of the models to check the model plausibility, the model variability and similarity through dimensionality reduction techniques; or some preliminary quantitative analysis of the conditioning accuracy or histogram distribution of property of interests. There is no detailed and enhanced evaluation workflow for the data conditioning performance and robust analysis of generated conditional realizations (Chan & Elsheikh, 2019; Dupont et al., 2018; Gao et al., 2020; Jo et al., 2020; Pan et al., 2022; Zhang et al., 2021)..

The enhanced model checking workflow, proposed by Liu et al. (2024) to support robust uncertainty around geological scenarios, utilizes the enhanced minimum acceptance checks for the quality of unconditional realizations generated by SinGAN. These checks include the experimental semivariograms, histograms, n-point histograms, pixel scale local distribution maps, and multiscale local distribution maps.

We propose the adoption of data conditioning checks for subsurface models generated with SinGAN and suggest a variety of metrics tailored to single image GAN methods to check and demonstrate the capabilities of SinGAN for data conditioning. Our conditioning checks include static small-scale local and large-scale exhaustive data conditioning through local accuracy checks characterized by the F score, the local uncertainty analysis through entropy relative to the number of conditioning locations, and the spatial nonstationarity reproduction checks. Our proposed checks include conditioning to geological concepts through multiscale spatial distribution, the number of connected geobodies checks, the spatial continuity checks through experimental semivariogram, and model facies proportion reproduction checks. We demonstrate the data conditioning workflow and the conditioning performance for one representative channelized deepwater reservoir model using image inpainting technique employed with SinGAN. We select the channelized geological model due to their representative and complex reservoir features common for deepwater and fluvial subsurface reservoirs and their geological concept conditioning performance is more straightforward to visualize and diagnose than other more chaotic depositional settings such as a delta. Our work shows a detailed diagnosis of data conditioning performance for geological subsurface models using SinGAN. We identify potential limitations in reproducing spatial connectivity, spatial nonstationarity and conditioning accuracy with different number of conditioning locations. Our proposed metrics and diagnostic approach provide guidance on the conditioning of single image GAN models for subsurface modeling and enhanced model conditioning checking essential for the application of these models to support uncertainty characterization and decision making.

The methodology section introduces the workflow of data conditioning of SinGAN for generating conditional geological facies models and a variety of conditional checks, followed by the summary quantification and conditioning performance analysis of generated conditional realizations in the results and discussions section. The conclusion section highlights the capability of data conditioning employed to single image GAN methods, potential limitations of data conditioning in GAN-related methods, and recommendations for future subsurface applications of deep learning for conditioning, such as the integration of our demonstrated conditional checks to the model to improve the conditioning quality.

## 2. Methodology

To demonstrate our proposed data conditioning checks for subsurface models generated with SinGAN, we first provide a workflow for data conditioning with SinGAN, followed by the introduction to a variety of conditioning checks tailored to single image GAN methods.

### 2.1 Workflow for data conditioning with SinGAN

We integrate data conditioning through post-optimization at the coarsest scale ($n = N$) of the pretrained SinGAN model. A random Gaussian noise vector $z_N$ is initiated at the coarsest scale, then the optimum noise vector, $z_{opt}$, is obtained by minimizing the object loss function (Equation 1) (Yeh et al., 2017) using a fast-convergence optimization method, COBYLA gradient-free optimizer (Powell, 2007).

$$\text{loss} = ||M \odot y - M \odot G(z_N)|| + \log(1 - D(G(z_N) * ||M - 1|| + M * y)) \qquad (1)$$

where M follows Dupont et al. (2018) and is set to be 5 in our case. D and G refer to the discriminator and generator in the pretrained SinGAN model, $z_N$ denotes a random Gaussian noise at the coarsest scale, and y refers to the original real image. Different from the original perceptual loss that only penalizes the generated realization $G(z)$, here we combine the generated realization and the real image to reduce the penalty on constrained realization solely and improve the flexibility of the optimization process. With an optimum random Gaussian noise $z_{opt}$ calculated from minimizing the loss function, the conditional realization, $G(z_{opt})$, is obtained through the pretrained SinGAN model.

2.2 Quality assessment of conditioning performance

Following are a variety of checks for the static small-scale local measures data conditioning, static large-scale exhaustive data conditioning, dynamic grade-forecasting data conditioning, and dynamic time-lapse data conditioning, considering the conditioning accuracy, conditioning uncertainty, subsurface model features reproduction, local scale, and multiscale spatial distribution reproduction.

To check SinGAN local data conditioning we apply the F1 score over the conditioning data locations, a commonly-used statistic to quantify classification accuracy in machine learning research and practice (Goutte & Gaussier, 2005; Sokolova et al., 2006). F1 score is a measure of predictive performance, which is the harmonic mean of precision and recall given the confusion metric (Rijsbergen, 1979) (Equation 2). Table 1 shows the confusion metric using the channelized model as an example. True positive (TP) refers to locations that are correctly reproduced as channels ("positive" here) in the generated realization, false positive (FP) refers to locations that are wrongly reproduced as channels in the generated realization, false negative (FN) means that locations are wrongly reproduced as mud ("negative" here), and true negative (TN) means that locations are correctly reproduced as mud. Then the precision and recall are calculated following Equation 3.

Table 1. Confusion metric of channelized models

|            |         | Truth   |     |
|------------|---------|---------|-----|
|            |         | channel | mud |
| Prediction | channel | TP      | FP  |
|            | mud     | FN      | TN  |

$$F1 = \frac{1}{\frac{1}{\text{precision}} + \frac{1}{\text{recall}}} \tag{2}$$

$$\text{precision} = \frac{TP}{TP + FP} \qquad \text{recall} = \frac{TP}{TP + FN} \tag{3}$$

Shannon entropy, a concept from information theory, is a measure of information, choice, and uncertainty. It quantifies the expected value of the information contained in a message (Shannon, 1948). This measure is crucial in determining the amount of uncertainty, or randomness inherent in a statistical mechanical system. The higher the entropy, the greater the uncertainty and the more information is required to describe the system (Ellis, 2012). We use Shannon entropy to quantify the local uncertainty over the SinGAN realization given conditioning data. The Shannon entropy map is calculated using Equation 4

$$H(x_{ij}) = -p(x_{ij})\log(p(x_{ij})) \tag{4}$$

where $p(x_{ij})$ is the probability of intensity value x at the pixel location, $(i, j)$, across all generated realizations, and $H(x_{ij})$ is the entropy value at this pixel location.

The multiscale local distribution checks the local, multiscale trend reproduction of spatial patterns by calculating the facies proportions within a moving window (Liu et al., 2024). This local facies proportions map is plotted as scatters of the facies proportions within the moving window of the generated realizations vs. the TI facies proportions with the same window size and location. The window size should be at a sufficient scale to smooth out local noise and calculate trends (Figure 1). The number of isolated components checks the connectedness of the geobodies in a subsurface model. Figure 2 shows a realization example with two isolated geobodies colored.

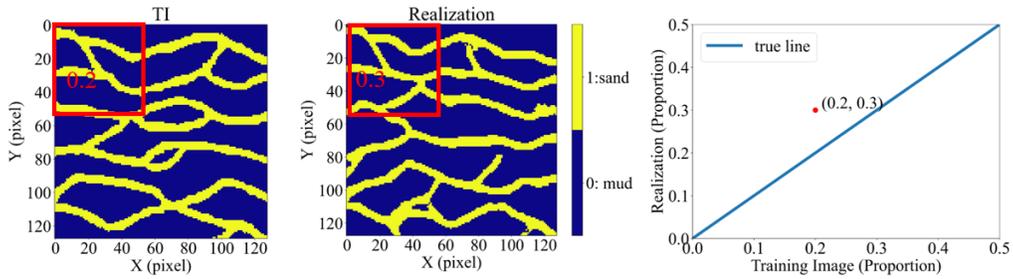

Figure 1. The illustration for multiscale local distribution checks. The channel proportions within a moving window are calculated of the TI (left) and the realization (middle), and the multiscale local distribution plot (right) is made by plotting the channel proportions of ground truth (TI) vs. the channel proportions of the realization.

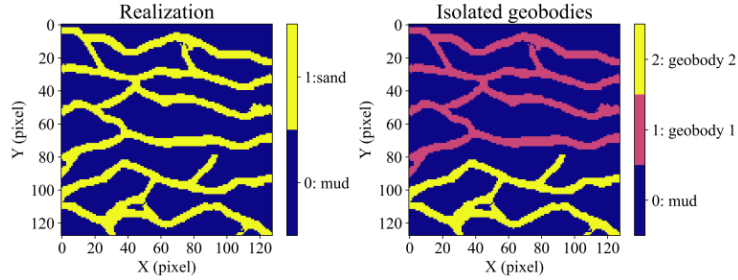

Figure 2. A channelized model realization example (left) and the associated isolated geobodies (right).

Subsurface model feature checks are the spatial continuity check quantified by the experimental semivariogram, the property of interests (such as facies proportions) reproduction check quantified by the histogram distributions, the local feature reproduction check quantified by the pixel scale local distribution maps. The indicator semivariogram, defined by the average squared difference of values separated by a lag distance (Equation 5), is calculated in major and minor directions and quantified by the statistics P25 and P75 for conditional cases and compared with that of the unconditional realizations.

$$\hat{\gamma}(\mathbf{h}) = \frac{1}{2N(\mathbf{h})} \sum_{\alpha=1}^{N(\mathbf{h})} [z(\mathbf{u}) - z(\mathbf{u} + \mathbf{h})]^2 \qquad (5)$$

where α represents $1, \ldots, N(\mathbf{h})$ pairs of values differences, $z(\mathbf{u}) - z(\mathbf{u} + \mathbf{h})$, at over lag distance $\mathbf{h}$.

The facies proportions histogram over multiple realizations is calculated and benchmarked with the baseline unconditional case to check if facies proportions can be reproduced correctly in data conditioning. The pixel scale local distribution maps summarize the average and variance pooled over realizations by-pixel and are compared with the unconditional case to check pixel-scale changes with conditioning integrated.

## 3. Results & Discussions

To check the conditioning performance of the workflow for data conditioning in SinGAN, we first train the SinGAN on a binary channelized model with cyclic sequences and calculate unconditional realizations. Cyclic sequences are vertical repetitive variations in the facies and petrophysical properties due to the fact of repetitively occurring geological phenomena over geological time (Pyrcz & Deutsch, 2014). Then we experiment with conditioning the trained SinGAN model on different number of random conditioning locations, N, where N = 10, 20, …, 80, and calculate 100 conditional realizations for each to evaluate conditioning performance in expectation and over the range of outcomes. We compare the quality check results of realizations conditioning to local well data with those of baseline unconditional realizations to evaluate the impact of local data conditioning. Figure 3 shows the training image (TI), an unconditional realization example, and two conditional realization examples, R1 and R2, of N = 10, 40, and 80 random conditioning well locations.

To check the conditioning accuracy, we calculate the F1 scores of 100 realizations conditioning to N well locations and baselined with the unconditional realizations at the same conditioning locations (Figure 4). The conditioning accuracy of the unconditional realizations is similar regardless of the number of conditioning with F1 score distributions confidence interval (CI) P25-P75 are around 0.6-0.8. As the number of conditioning increases, the F1 score distribution starts to decrease for the conditional realizations, indicating that the conditioning is more difficult to achieve, specifically at the 80 conditioning data.

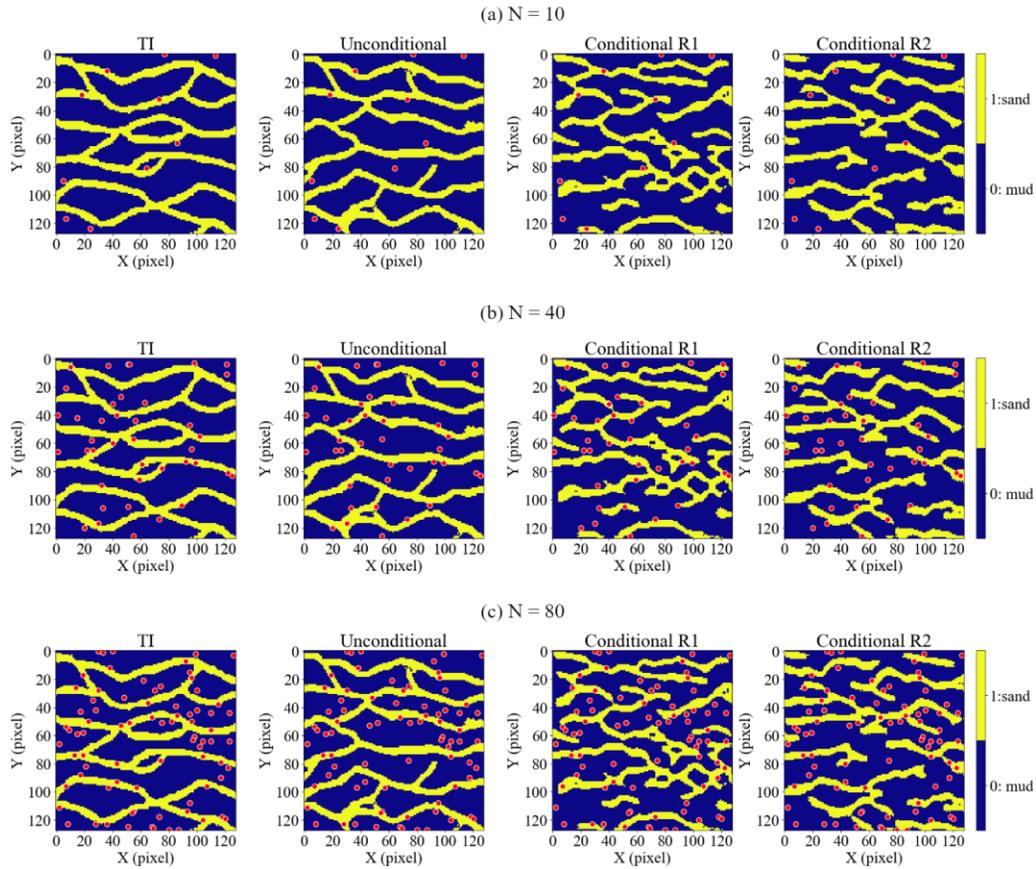

Figure 3. The channelized model training image, unconditional realization example, and two conditional realization examples for (a) 10 points, (b) 40 points, and (c) 80 points with conditioning locations scattered in red.

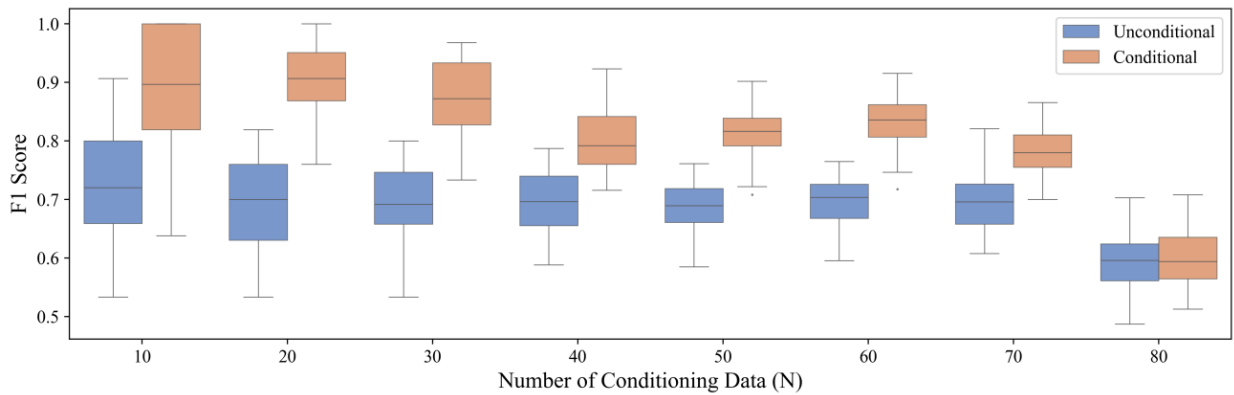

Figure 4. The box plots of F1 score distributions of conditional realizations and unconditional realizations regarding different number of conditioning data. The number of data locations ranges from 10 to 80 with a step of 10.

To check the facies uncertainty reproduction at constrained locations, Figure 5 presents the entropy maps of unconditional realizations and conditional realizations for different number of conditioning data. Compared to the unconditional entropy map with higher entropy values at the central region, the conditional entropy maps show lower entropy values specifically at these constrained locations, indicating that the conditioning process decreases the uncertainty of facies reproduction to some degree at constrained locations. To visualize and evaluate the global reproduction of uncertainty with the increasing number of conditioning data, we compare the histograms of entropy values at constrained locations for 50 random unconditional realizations and conditional realizations, respectively, shown in Figure 6. The histograms show that entropy values at constrained locations for conditional cases are generally smaller than those of the unconditional case. However, the entropy value distributions are similar to that of the unconditional case with an increasing number of data, specifically at 80 data, implying the difficulty of conditioning with an increasing number of data.

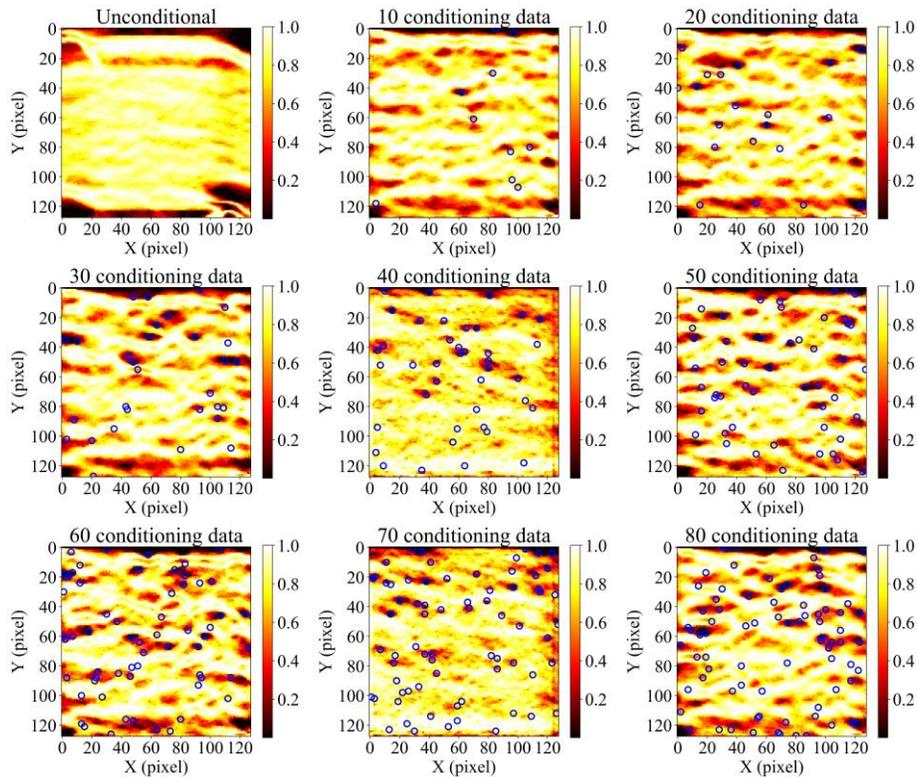

Figure 5. The entropy maps of the unconditional realizations and conditional realizations for different number of conditioning data depicted in blue circles.

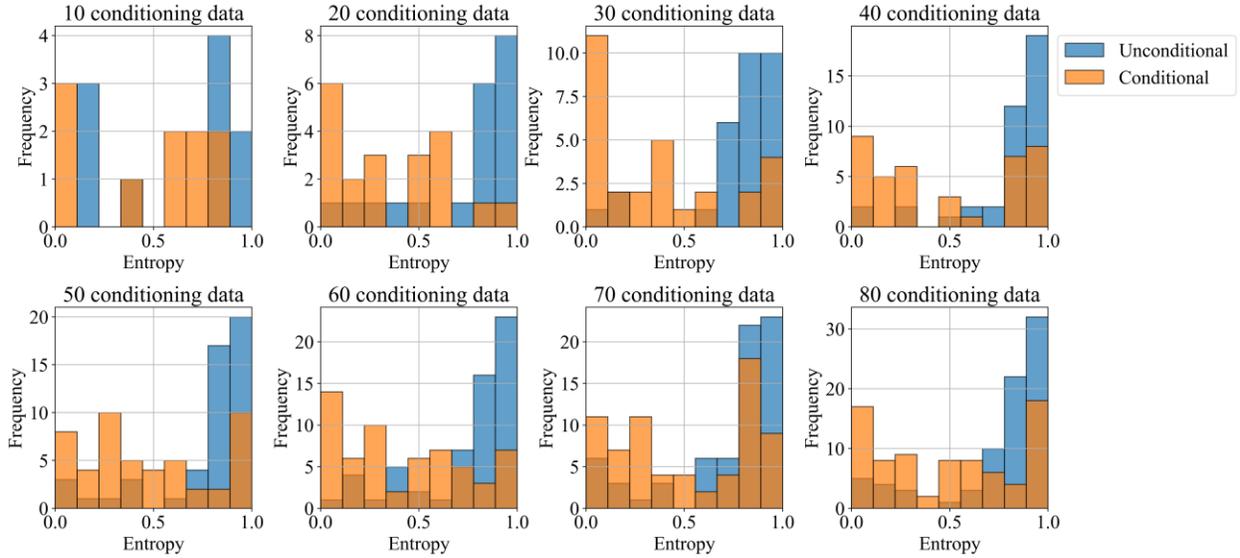

Figure 6. The histograms of entropy values at the constrained locations for unconditional case (blue bars) and conditional cases (orange bars) of different number of conditioning data.

We calculate the by-lag experimental semivariograms of major and minor directions and plot the confidence intervals (P10-P90) to visualize the spatial continuity reproduction (Figure 7 and 8 are for major and minor directions, respectively). Overall, the major and minor spatial continuity can be reproduced in conditional realizations regardless of number of conditioning data.

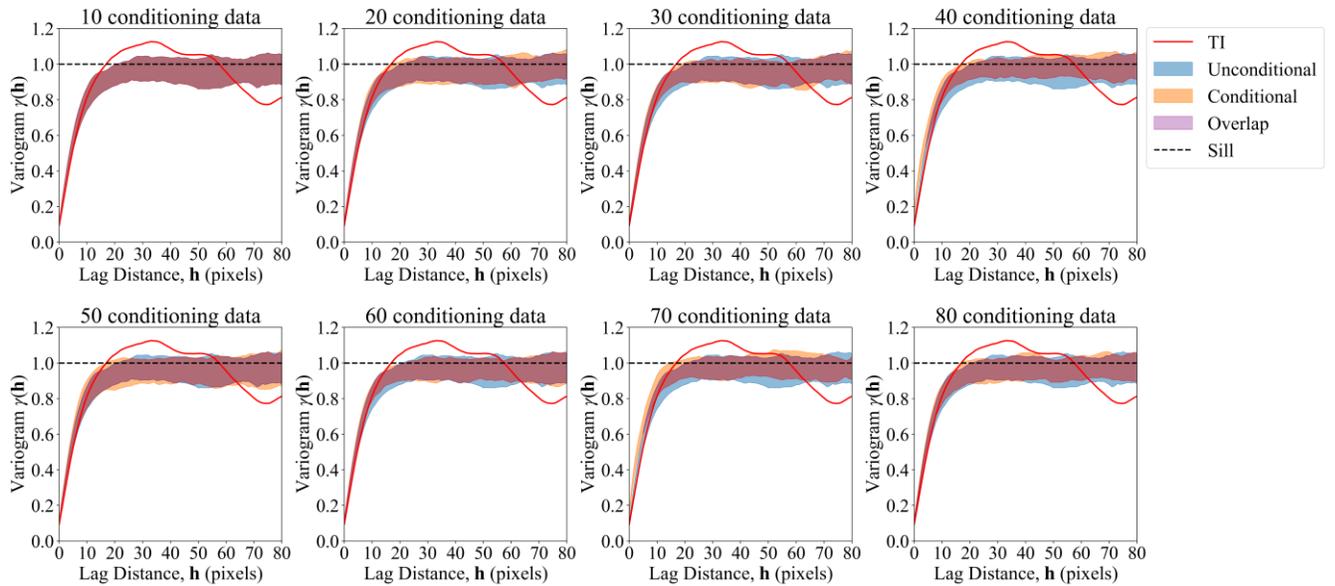

Figure 7. The by-lag major experimental semivariograms confidence interval (CI): P10-P90 for unconditional realizations and conditional realizations of different number of conditioning data. The TI is in red, the unconditional case is in blue, and the conditional case is in orange. The overlap area is depicted in purple.

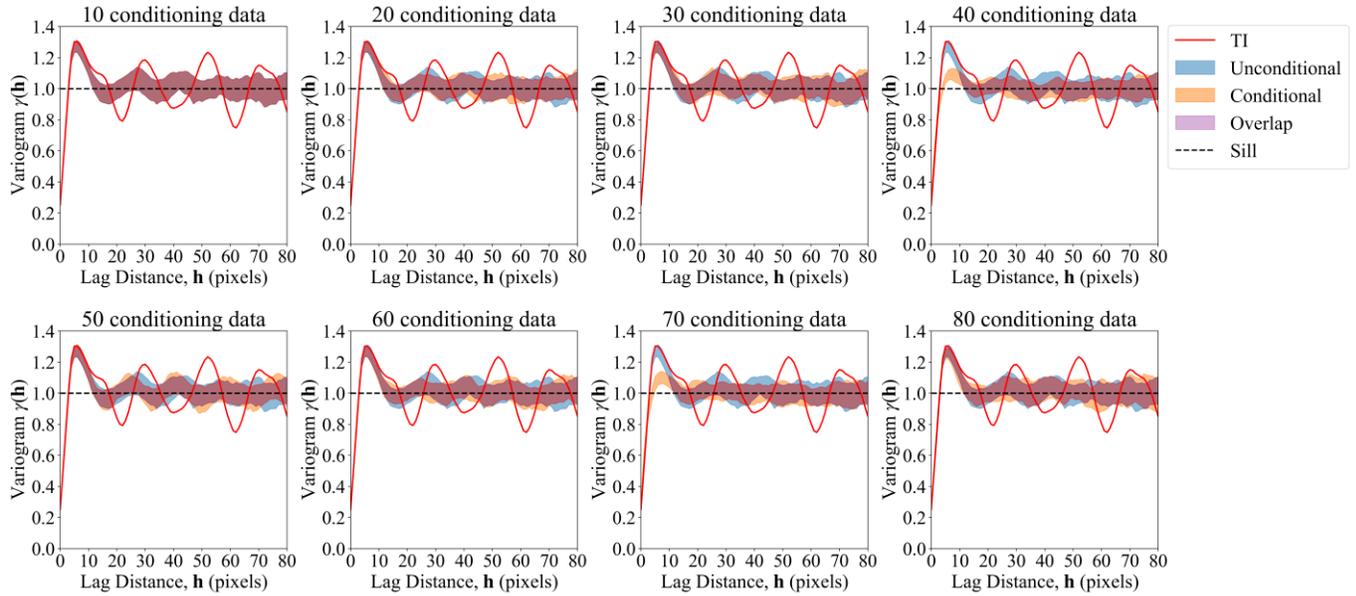

Figure 8. The by-lag minor experimental semivariograms confidence interval (CI): P10-P90 for unconditional realizations and conditional realizations of different number of conditioning data. The TI is in red, the unconditional case is in blue, and the conditional case is in orange. The overlap area is depicted in purple.

We calculate the channel proportion histograms to check the global facies proportions reproduction in conditional realizations by comparing them with the channel proportions of unconditional realizations and TI in Figure 9. Overall the channel proportion distributions are close to the channel proportion of the TI and overlap the channel proportion distributions of the unconditional case except for 40 and 70 conditioning data, with a slight distribution shift of around 0.02 (7% of the channel proportion of TI).

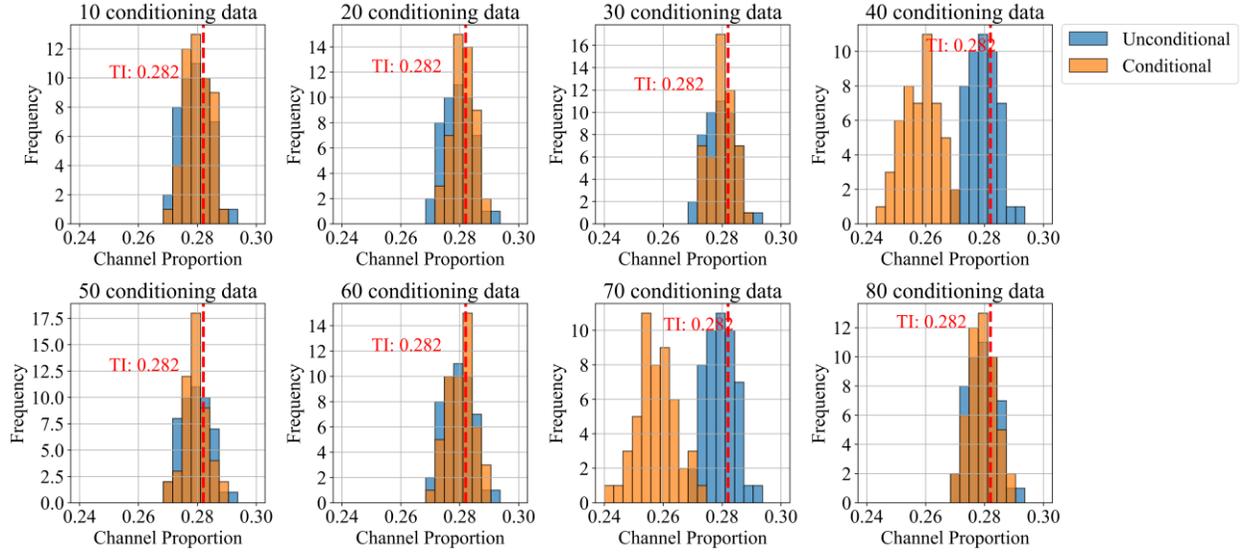

Figure 9. The histograms of channel proportions of generated realizations for unconditional case (blue bars) and conditioning cases (orange bars) with different number of conditioning data. The TI channel proportion is shown as a red dashed line.

The local average maps (Figure 10) and dispersion maps (Figure 11) show the pixel scale local distribution maps calculated over all realizations for unconditional case and each conditional case. Compared to the local average map of the unconditional case that shows high channel reproduction at edges (very high and very low local channel proportions over all realizations), the average maps of the conditional case show that the channels are prone to reproduce along specific patterns, especially around the conditioning locations. We also observe that the variety of channel reproduction of unconditional case mainly locate in the central region of the image, and this variety is becoming more restricted in conditional cases (less yellow colored region).

We calculate the channel proportions within a moving window with a stride of 1 to check the local, multiscale trend reproduction of the spatial distributions. We show the confidence interval of channel proportions rather than all scatters here for more clear visualization in Figure 12. Compared with the confidence interval of the unconditional case that aligns well with the true line, the confidence interval of moving window channel proportions is more redundant for conditional cases, especially at 40 and 70 conditioning data that show around 0.02 smaller channel proportion reproduction (also reflected in the channel proportions histogram distributions). This implies the possible less channel reproduction and local trend production in the generated conditional realizations.

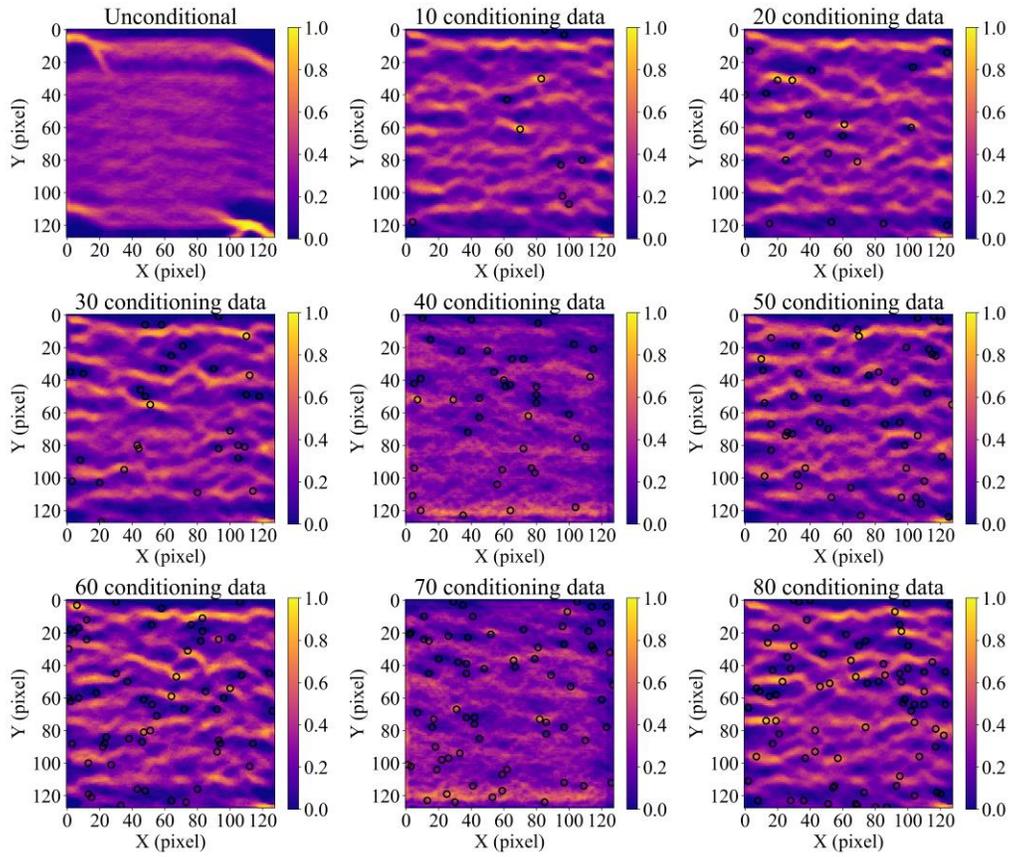

Figure 10. The local average maps of generated realizations for unconditional case and each conditional case with conditioning locations depicted in black.

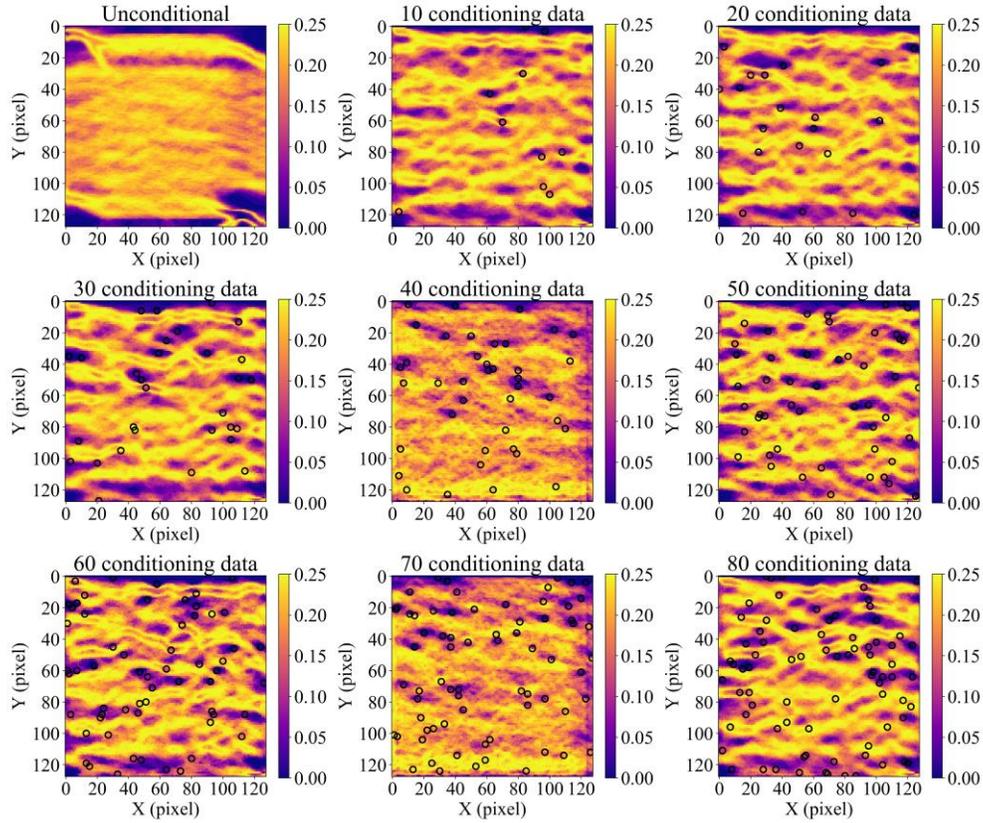

Figure 11. The local dispersion maps of generated realizations for unconditional case and each conditional case with conditioning locations depicted in black.

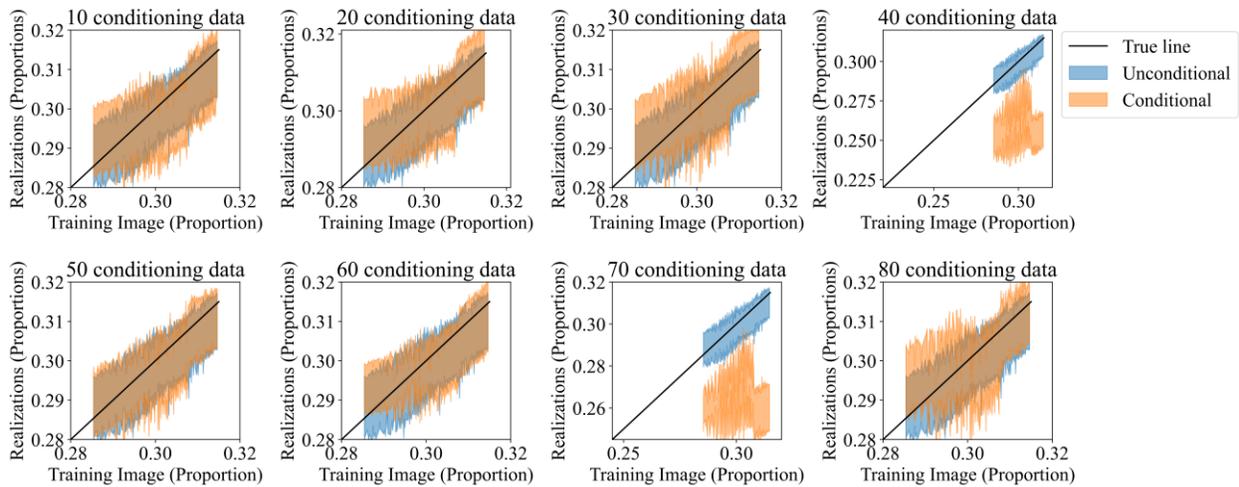

Figure 12. Confidence intervals (P25-P75) of channel proportions within the moving window of size 110 for each conditional case. The confidence interval of the baseline unconditional case is depicted in blue and the conditional case is in orange.

We calculate the number of isolated channels of generated realizations for conditional cases and compare them with the unconditional case to check the impacts of conditioning on geobodies connectivity. Figure 13 shows TI, realization examples of unconditional case and different conditional cases. The TI has three disconnected geobodies and the channels become more disconnected for data conditioning realizations. We calculate the number of disconnected geobodies for each conditioning case and compare the distributions with the unconditional case, shown in Figure 14. We observe that the number of channels for unconditional case are close to that of the TI, however, the number of geobodies increases to the average of around 10 for conditional cases, indicating the disconnectedness problem in generated realizations with data conditioning.

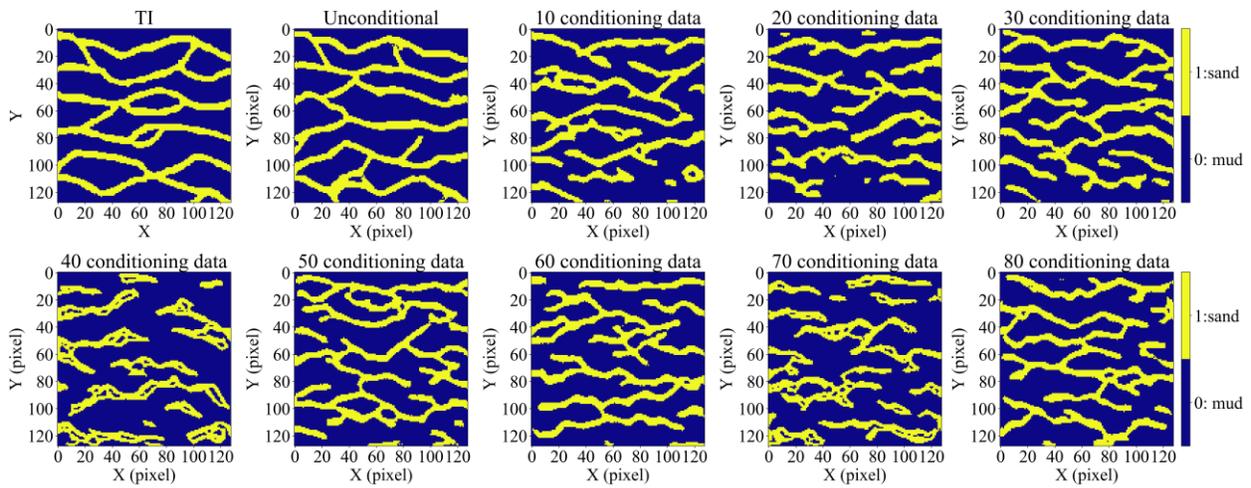

Figure 13. The TI and realization examples for unconditional case, and conditional cases of 10, 20…, 80 conditioning data.

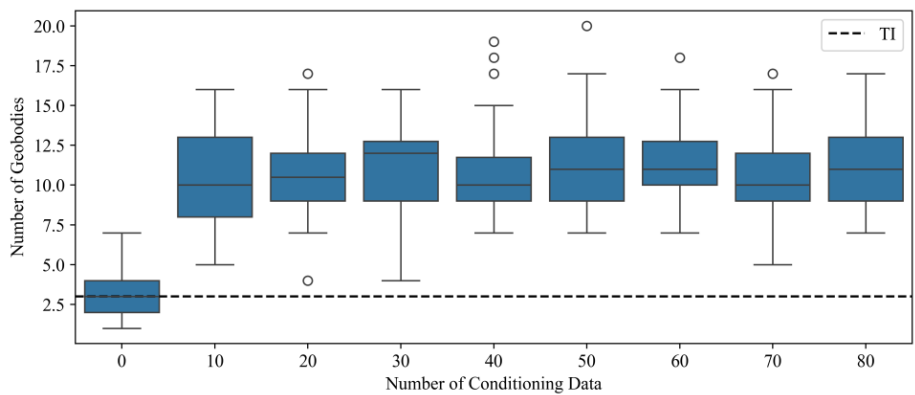

Figure 14. The box plots of the number of disconnected geobodies for unconditional realizations, and conditional realizations for different number of conditioning data. The number of disconnected geobodies of TI is 3 with a black dashed line depicted.

The above checks show that realizations through data conditioning with SinGAN can reproduce the model patterns and spatial continuity with observed limits related to exact data reproduction, local trend reproduction, facies proportions reproduction, geobody connectivity. The averages of some metrics results are summarized in Table 1. There is a limitation of conditioning accuracy reflected from decreasing F1 score with increasing data conditioning. The global channel proportion distributions may be biased. Meanwhile, more data conditioning will likely break out the connectivity of geobodies.

Table 1. Summary results for averages of the proposed metrics with the most impacts.

| Checks | Number of Data Conditioning | | | | | | | | |
| --- | --- | --- | --- | --- | --- | --- | --- | --- | --- |
| | TI | 10 | 20 | 30 | 40 | 50 | 60 | 70 | 80 |
| F1 Score | - | 0.890 | 0.908 | 0.871 | 0.802 | 0.813 | 0.833 | 0.782 | 0.600 |
| Channel Proportions Distribution | 0.282 | 0.280 | 0.281 | 0.280 | **0.258** | 0.279 | 0.280 | **0.258** | 0.279 |
| Number of Geobodies | 3 | **10.44** | **10.60** | **10.96** | **10.50** | **11.44** | **11.32** | **10.6** | **11.34** |

4. Conclusions

The data conditioning for subsurface modeling with SinGAN is demonstrated and checked with our proposed single training image generative AI realization minimum acceptance criteria. These checks include local conditioning data accuracy through F1 score, local conditioning uncertainty through Shannon entropy, spatial continuity through experimental semivariogram, local scale, and multiscale spatial trend relationships reproduction, the global facies proportion histograms, and the geobodies connectivity checks. These checks are conducted on conditional realizations of different cases and baselined with the unconditional realizations. We perform data conditioning using the proposed Equation 1. However, we notice the limitation of this conditioning accuracy as the number of conditioning data is increased. Additionally, although the spatial continuity and facies proportion distributions can be reproduced, we observe there are pitfalls related to accurate data conditioning reproduction, through a decrease in the variety of generated conditional realizations, and multiscale spatial relationship reproduction. This conditioning process can also cause unstable channel realization in terms of channel connectivity reflected by increasing disconnected geobodies. Future work can be focused on applying these minimum acceptance conditioning checks criteria on other generative models to improve the data conditioning performance.

## 5. Acknowledgments

The authors thank the support from the Digital Reservoir Characterization Technology (DIRECT) Industry Affiliate Program at the University of Texas at Austin.

## Appendix 1. SinGAN

The schematic architecture of SinGAN is shown in Figure 1. The SinGAN is a pyramid of generators and discriminators both sharing the same architecture, five convolutional layers block (Figure 2), at each scale. At the coarsest scale (n = 0), the generator only accepts a random Gaussian noise and outputs a fake realization. For the remaining scales, the generators accept a random Gaussian noise and an upsampled realization from the previous coarse scale. The discriminators discriminate the fake realization (x) with the real image (y) at each scale. The conditioning of SinGAN is conducted after the unconditional training of SinGAN from Liu et al. (2024).

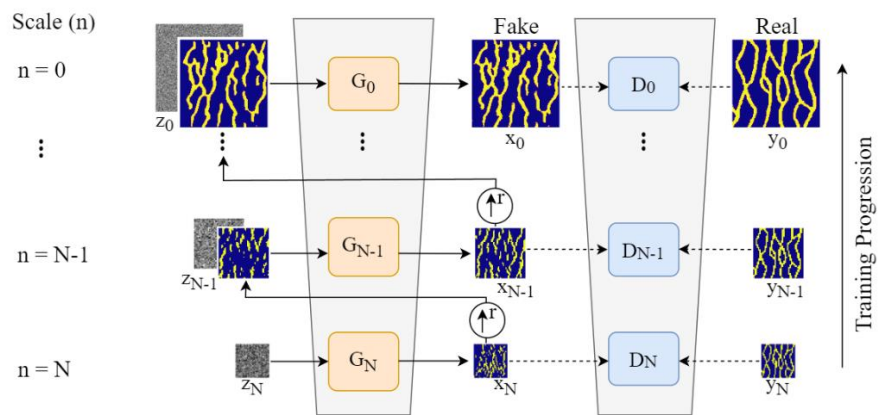

Figure 1. SinGAN architecture for geological facies modeling (Liu et al. (2024)). The architecture has a pyramid of generator and discriminators, both sharing the same architecture from coarse scale (n = N) to fine scale (n = 0).

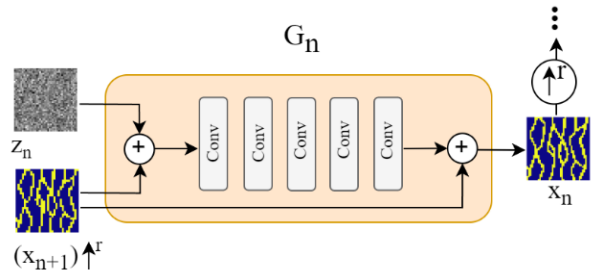

Figure 2. The generator architecture in SinGAN (Liu et al. (2024)). At scale $n$ ($n \leq N - 1$), the generator receives a Gaussian random noise with an upsampled realization of the previous scale and feeds them to five convolutional layers, and outputs a fake realization.